\documentclass[conference]{IEEEtran}
\IEEEoverridecommandlockouts

\usepackage{url}
\usepackage{booktabs}

\usepackage{stfloats}
\usepackage{placeins}

\usepackage{cite}
\usepackage{amsmath,amssymb,amsfonts}
\usepackage{algorithmic}
\usepackage{graphicx}
\usepackage{textcomp}
\usepackage{xcolor}
\def\BibTeX{{\rm B\kern-.05em{\sc i\kern-.025em b}\kern-.08em
    T\kern-.1667em\lower.7ex\hbox{E}\kern-.125emX}}
\begin{document}

\title{Forecasting Success of Computer Science Professors and Students Based on Their Academic and Personal Backgrounds}

\author{\IEEEauthorblockN{1\textsuperscript{st} Ghazal Kalhor}
\IEEEauthorblockA{\textit{School of Electrical and Computer Engineering} \\
\textit{University of Tehran)}\\
Tehran, Iran \\
kalhor.ghazal@ut.ac.ir}
\and
\IEEEauthorblockN{2\textsuperscript{nd} Behnam Bahrak}
\IEEEauthorblockA{
\textit{Tehran Institute for Advanced Studies}\\
Tehran, Iran \\
b.bahrak@teias.institute}
}

\maketitle

\begin{abstract}
After completing their undergraduate studies, many computer science (CS) students apply for competitive graduate programs in North America. Their long-term goal is often to be hired by one of the big five tech companies or to become a faculty member. Therefore, being aware of the role of admission criteria may help them choose the best path towards their goals. In this paper, we analyze the influence of students' previous universities on their chances of being accepted to prestigious North American universities and returning to academia as professors in the future. Our findings demonstrate that the ranking of their prior universities is a significant factor in achieving their goals. We then illustrate that there is a bias in the undergraduate institutions of students admitted to the top 25 computer science programs. Finally, we employ machine learning models to forecast the success of professors at these universities. We achieved an RMSE of 7.85 for this prediction task.
\end{abstract}

\begin{IEEEkeywords}
graduate admission, academia, educational background, computer science, machine learning, success prediction
\end{IEEEkeywords}

\section{Introduction}
North America is a popular destination for students from around the world who aspire to pursue their graduate studies \cite{she2013international, ghai2015analysis}. According to Slonim et al. \cite{slonim2008outlook}, there has been a growing trend in computer science graduate enrollments at American and Canadian universities. This trend can be attributed to various factors. Firstly, there is an increasing interest among students in the field of artificial intelligence within computer science \cite{hendler2008avoiding}. Additionally, computer science graduates have access to numerous job opportunities across different industries \cite{stepanova2021hiring}. Moreover, the competitive salaries and job security associated with careers in computer science contribute to this trend \cite{sade2019factors}. However, obtaining a master's or doctoral degree from a university in the USA or Canada is often a requirement for finding professional jobs in these countries \cite{wendler2012pathways}.

Several previous studies have examined the criteria for graduate admissions. In our previous study \cite{kalhor2023diversity}, we analyzed the impact of students' personal background, such as gender and home country, on their admissions to computer science graduate programs at top North American universities. We found evidence of a nationality bias in the admission process, but no gender bias was observed. Turner and Hudson \cite{newkirkturner2022do} identified unintended bias in reference letters for Black applicants and proposed suggestions for improvement. Additionally, a student's statement of purpose plays a vital role in the application process for admission to elite universities \cite{kanojia2017statement}.

Furthermore, students' educational background, including their grade point average (GPA), research experience, and previous universities, significantly impact their chances of admission to top-tier universities. High-ranking universities often set minimum GPA requirements and reject applicants who fall below the specified cut-off \cite{raghunathan2010demystifying}. Moreover, having publications in prestigious venues is an indicator of research potential, particularly for doctoral admissions \cite{michel2019graduate}. These criteria are explicitly stated in universities' application instructions. However, the selectivity of a student's prior university is an implicit factor that admissions committees consider when reviewing applications and making admission decisions \cite{posselt2014toward}.

Despite the importance of undergraduate universities in the admissions process, only a few studies have analyzed their role in gaining admission to prestigious graduate programs. Lang \cite{lang1987stratification} used regression techniques to assess the impact of student characteristics, such as gender, race, and the rank of their bachelor's institution, on predicting the rank of their graduate university. The study demonstrated that the rank of a student's undergraduate institution is the most influential factor in predicting attendance at leading universities.  Zhang \cite{zhang2005advance} investigated how the quality of a student's college influences the quality of graduate programs they are admitted to. The research confirmed that the quality of a student's college has a significant positive impact on their chances of enrolling in graduate programs in the United States.

Some studies have attempted to predict faculty members' success in terms of their h-index. McCarty et al. \cite{McCarty2013predicting} predicted authors' h-index based on the characteristics of their co-author networks, finding that having more co-authors results in higher h-indexes. Ayaz \cite{ayaz2018predicting} and Ibanez et al. \cite{ibanez2011predicting} forecasted professors' h-index over time, suggesting that predictions for shorter periods are more accurate.

Building upon the previous studies, our research examines the influence of students' previous universities on their success in the admission process for prestigious universities. Additionally, we explore the prior universities of faculty members and compare them with those of their mentees to uncover any biases. We also employ regression techniques to predict professors' h-index based on various factors. For our analysis, we utilize the Advisor Student Data that we introduced in \cite{kalhor2023diversity}. It is important to note that the information related to prior universities in this dataset is crucial but understudied.

Our contributions can be summarized as follows:

\begin{enumerate}
\item We analyze the distributions of the previous universities/colleges students and advisors have attended.
\item We assess the correlation between the ranking of students’ prior universities and their chances of getting accepted to high-ranking universities.
\item We examine whether professors at top universities obtained their doctoral degrees from these institutions.
\item We conduct a proportion hypothesis test to determine if advisors are more likely to admit students from their own bachelor's universities.
\item We construct a network of student-advisor prior universities and calculate various metrics to investigate it.
\item We compare the admissions of female and male students in both the PhD and direct PhD programs.
\item We predict professors' h-index based on their academic and personal backgrounds.
\end{enumerate}

\section{Results and discussion}\label{sec:sec3}
In this section, we outline the steps taken to address our research questions. We subsequently offer insightful interpretations of our findings.

\subsection{Previous universities distributions}
\begin{figure}[!ht]
\centering
\includegraphics[width=9cm, keepaspectratio]{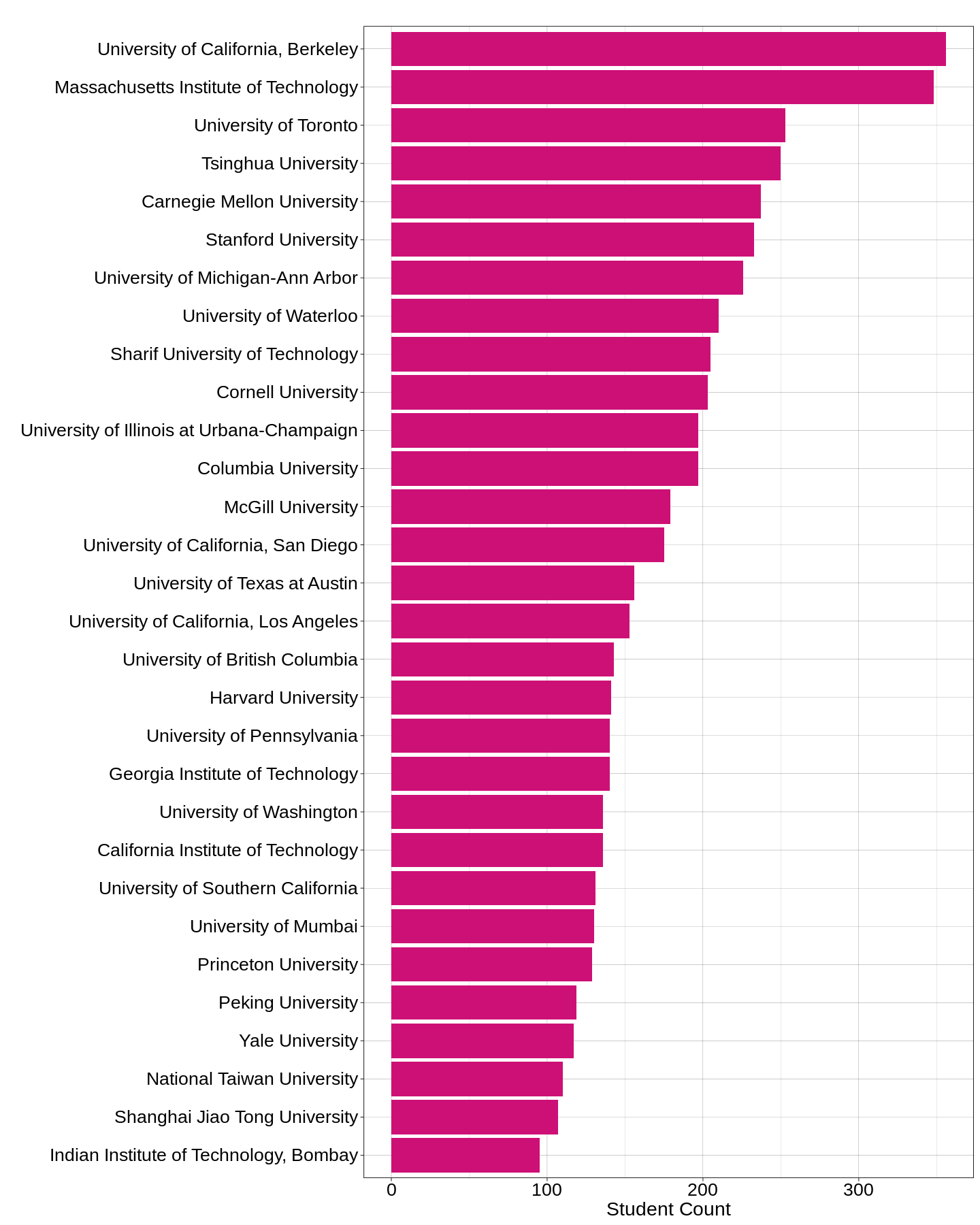}
\caption{Distribution of students' prior universities in the dataset.}
\label{fig:fig2}
\end{figure}

\begin{figure}[!ht]
\centering
\includegraphics[width=9cm, keepaspectratio]{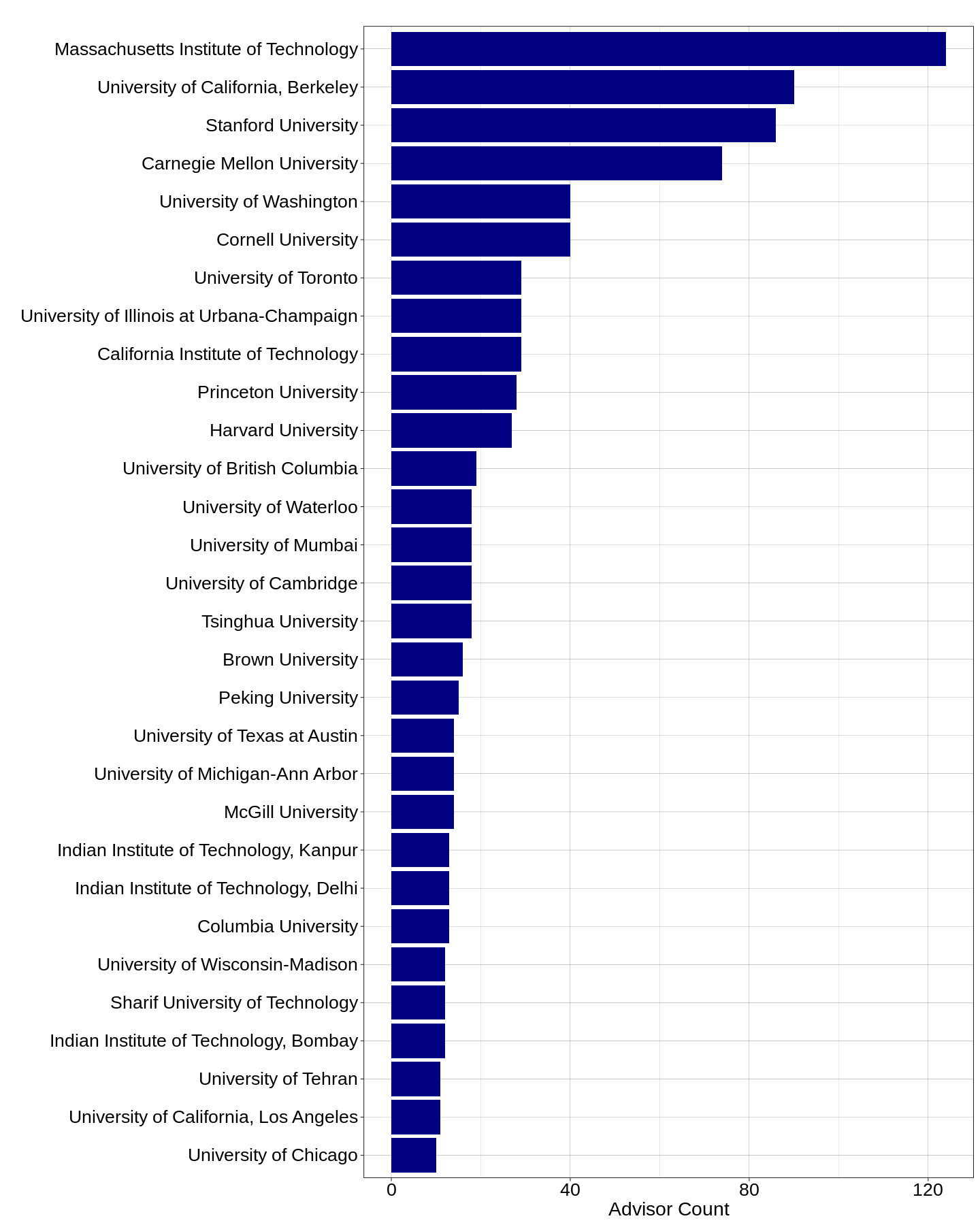}
\caption{Distribution of advisors' prior universities in the dataset.}
\label{fig:fig3}
\end{figure}

In this part, we examine the distribution of universities where students and advisors have previously graduated from. Figures \ref{fig:fig2} and \ref{fig:fig3} display the prevalence of the most commonly represented universities among students and faculty members, respectively. According to Figure \ref{fig:fig2}, a significant proportion of students admitted to the top 25 universities have obtained their degrees from esteemed global computer science institutions. Furthermore, Figure \ref{fig:fig3} illustrates that professors predominantly hold at least one degree in computer science from the top 25 universities in North America. We further analyze and discuss these observations in subsequent parts of the paper.

\subsection{Relationship between acceptance and ranking}

In this part, we examine the relationship between admissions and the ranking of students' previous universities. To accomplish this, we sort the previous universities in descending order based on the number of students in the dataset who graduated from them with a bachelor's or master's degree. Subsequently, we provide a ranking for the top 30 universities using this order and compare it with the QS World University Ranking for Computer Science 2022\footnote{\url{https://www.topuniversities.com/university-rankings/university-subject-rankings/2022/computer-science-information-systems}} by calculating Kendall's tau \cite{kendall1938new}. The correlation between the two rankings is 0.4244, with a p-value of 0.0010. Based on these findings, we conclude that there is a relatively strong positive association between the ranking of previous universities and the number of students admitted to high-ranking North American universities. In other words, universities with higher rankings are more likely to have a greater number of students accepted into top North American institutions. Therefore, in line with previous studies \cite{lang1987stratification, zhang2005advance}, we establish that the ranking of students' prior universities is a significant factor in graduate admissions. 

\subsection{Evaluating faculty members’ PhD university impact}

In this part, our objective is to determine if a majority of the advisors in the dataset obtained their doctoral degrees from one of the top 25 ranked North American universities. To accomplish this, we calculate the proportion of faculty members who graduated from these universities with a PhD degree. Since the conditions of independence and success-failure are met, we can assume that the sampling distribution of the proportion is approximately normal, with a mean of 0.5 and a standard deviation of 0.0184. Furthermore, our point estimate (the observed ratio) is 0.7395. Therefore, our hypothesis test is formulated as follows:

\begin{eqnarray}
    \begin{gathered}
        H_0: p_{top \: ranking \: university \: ratio} = 0.5 \\
        H_a: p_{top \: ranking \: university \: ratio} > 0.5
    \end{gathered}
\end{eqnarray}

Using a z-test, we obtained a p-value of 3.5639e-39, which is significantly lower than our chosen significance level of 0.05. Therefore, we reject the null hypothesis. This provides strong evidence to support the conclusion that the majority of computer science faculty members at high-ranking North American universities hold a PhD degree from one of these institutions. This finding is consistent with Crane's research \cite{crane1970academic}, which demonstrated a higher likelihood of PhD graduates from prestigious departments being hired as professors at top universities. Furthermore, Bedeian et al. \cite{bedeian2010doctoral} stated that the reputation and prestige of a student's doctoral university directly impact the perceived quality and reputation of the academic appointment they receive.

\subsection{Assessing bachelor’s university bias}

In this part, our aim is to test whether advisors demonstrate a preference for accepting students who graduated from their own previous universities. To achieve this objective, we employ a simulation-based approach similar to the methodology used to assess nationality bias in \cite{kalhor2023diversity}. However, since some individuals may have attended more than one university, we need to preprocess the dataset to convert the columns representing students' and advisors' prior universities into single-value representations. Specifically, we focus solely on the universities from which students and advisors obtained their bachelor's degrees. It is worth noting that undergraduate institutions are a better indicator as many students directly apply to PhD programs with a bachelor's degree.

\begin{figure}[!ht]
\centering
\includegraphics[width=9cm, keepaspectratio]{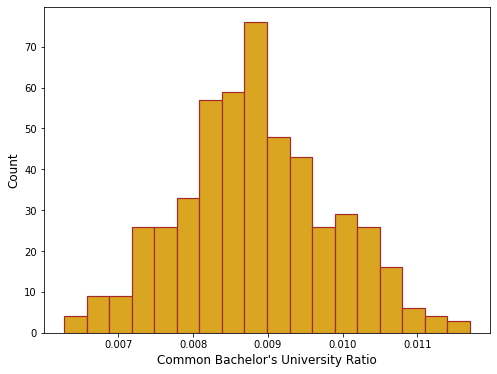}
\caption{Histogram of advisor-student common bachelor's university ratio. The number of simulations is 500.}
\label{fig:fig4}
\end{figure}

In each of the 500 iterations, we generate 9,400 advisor-student pairs, where the prior university for each component is selected with a probability equivalent to the observed proportion in the dataset. This simulation yields a distribution that approximates a normal distribution, with a mean of 0.0088 and a standard deviation of 0.0042, as depicted in Figure \ref{fig:fig4}. The ratio of advisor-student pairs sharing the same bachelor's universities in the dataset is 0.0289. Consequently, we need to assess the likelihood of observing such a value within the simulated distribution. Therefore, our hypothesis test is formulated as follows:

\begin{eqnarray}
    \begin{gathered}
        H_0: p_{common \: bachelor's \: university \: ratio} = 0.0088 \\
        H_a: p_{common \: bachelor's \: university \: ratio} \neq 0.0088
    \end{gathered}
\end{eqnarray}

Employing a z-test, we obtain a p-value of 1.5427e-6, which is lower than our significance level of 0.05. Therefore, we reject the null hypothesis and conclude that the data provide strong evidence of a preference for the bachelor's university in the acceptance of graduate students.

\subsection{Exploring prior universities network}

\begin{figure}[!ht]
\centering
\includegraphics[width=8.5cm, keepaspectratio]{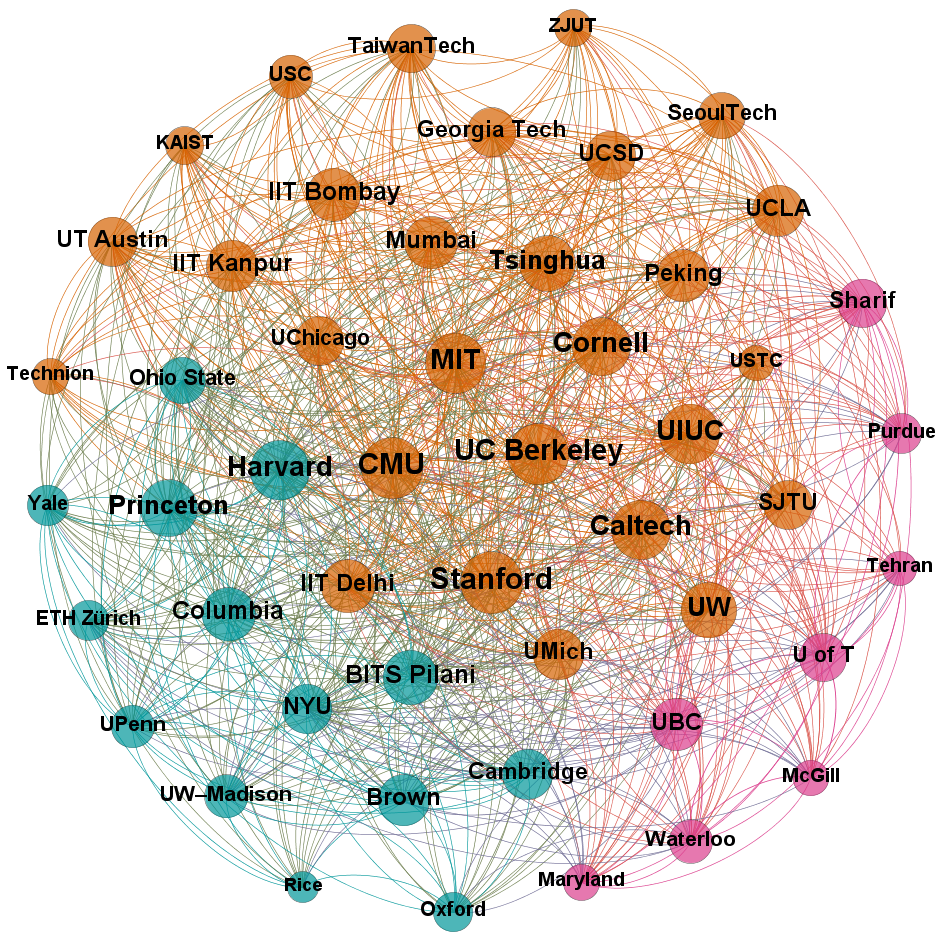}
\caption{A subgraph of the prior universities network. The color of nodes represents their respective community. The size of nodes corresponds to their authority value, while the size of their label corresponds to their closeness centrality. The thickness of edges indicates their weight.}
\label{fig:fig5}
\end{figure}

We create an undirected weighted network that depicts the relationships between students and advisors based on their respective previous universities. In this network, each node represents a prior university of either students or faculty members. An edge between universities $a$ and $b$, with a weight of $W$, indicates the presence of $W$ student-advisor relations between students from university $a$ and advisors from university $b$. Figure \ref{fig:fig5} presents a subgraph of this network, focusing on the first 50 universities with the highest node degrees. We then apply the disparity filter algorithm \cite{serrano2009extracting} to remove insignificant ties from the network. Additionally, we identify the communities within the network using the Louvain community detection algorithm \cite{blondel2008fast}, assigning a specific color to each node based on its community.

As observed, Canadian and Iranian universities are found in the same community. This result aligns with the fact that many Iranians choose to apply to Canadian universities due to the streamlined process of becoming a permanent resident in Canada, which is more efficient compared to the USA \cite{woroby2015immigration}. Moreover, Canada offers more flexible visa and immigration policies for international students \cite{she2013international}. The orange community predominantly comprises Chinese, Indian, and South Korean universities, reflecting the expected high number of international students from these countries studying at American universities \cite{tshibaka2018understanding}. The third community consists of European universities, providing further evidence of a nationality bias in student admissions.

Stanford University holds the highest values for both authority \cite{kleinberg1999authoritative} and closeness centrality \cite{freeman2002centrality}, indicating its commitment to promoting diversity and inclusivity\footnote{See \url{https://ideal.stanford.edu/about-ideal/diversity-statement}}.

\subsection{Analyzing doctorate admissions}

\begin{figure}[!ht]
\centering
\includegraphics[width=7cm, keepaspectratio]{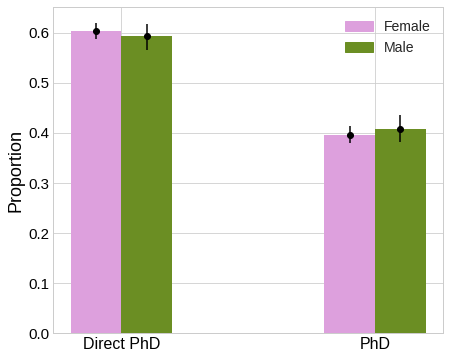}
\caption{Gender-disaggregated bar plot of PhD and direct PhD students. 95\% confidence intervals shown, with standard errors calculated using bootstrapping \cite{diciccio1996bootstrap}.}
\label{fig:fig6}
\end{figure}

In this part, we examine the distribution of PhD and direct PhD admissions among female and male students. Figure \ref{fig:fig6} illustrates that both genders have a higher proportion of students being admitted to doctoral programs without obtaining a master's degree. This observation aligns with the fact that having a master's degree is not a requirement for acceptance to computer science doctoral programs at top North American universities. Furthermore, the limited availability of full funding positions for master's students contributes to the preference of applying directly to PhD programs \cite{pottinger1999choosing}. Additionally, the ratio of direct PhD admissions to PhD admissions is higher for women compared to men. In other words, female students are more likely to be accepted into PhD programs directly after completing their bachelor's degree, in comparison to their male counterparts.

\subsection{Predicting professors' success}

In this part, we employ various machine learning techniques to assess the impact of professors' academic and personal backgrounds on their success. Our success metric is the h-index, and we utilize a range of features, including academic rank, gender, home country, research fields, previous universities, citation count, publication count, first paper year, and the nationality entropy of their research groups as predictive factors. Since machine learning models require numerical input, we convert categorical features such as home country and research fields into a one-hot representation. Additionally, we remove columns with more than 95\% zero values to ensure that our predictors are not excessively sparse, resulting in a final set of 28 columns.

We randomly allocate 80\% of the rows for training, reserving the remaining rows for the test set. To determine the optimal parameters for each model, we utilize the grid-search technique provided by the scikit-learn library \cite{pedregosa2011scikit}. Furthermore, we implement 5-fold cross-validation to mitigate the risk of overfitting in our models. Table \ref{tab:tab2} provides an overview of the evaluation metrics for forecasting faculty members' h-index. As shown, the neural network outperforms other regression models across most of the metrics. In our best neural network, the activation function of the hidden layers is ReLU, alpha is set at 0.001, the initial learning rate is 0.001, and the batch size is 100. The lowest RMSE achieved for this task is 7.89, with the highest adjusted R-squared of 0.88.

\begin{table*}[!t]
\centering
  \caption{Evaluation metrics of faculty members' h-index prediction regressors.}
  \label{tab:tab2}
  \begin{tabular}{cccccccl}
    \toprule
    Regressor &$R^2$ & Adj. $R^2$ & MAE & RMSLE & MSE & RMSE\\
    \midrule
    Linear Regression & 0.83 & 0.80 & 7.88 & 0.34 & 104.06 & 10.20\\
    
    Bayesian Ridge & 0.84 & 0.80 & 7.69 & 0.33 & 99.79 & 9.99  \\
    
    Lasso Regression & 0.84 & 0.81 & 7.71 & 0.33 & 98.51 & 9.93  \\

    Ridge Regression & 0.84 & 0.81 & 7.66 & 0.33 & 99.35 & 9.97  \\
    
    Decision Tree & 0.82 & 0.78 & 6.71 & 0.22 & 111.67 & 10.57  \\
    
    Random Forest & 0.89 & 0.87 & \textbf{5.38} & \textbf{0.18} & 68.23 & 8.26  \\
    
    K-Nearest Neighbors & 0.78 & 0.73 & 7.30 & 0.24 & 138.36 & 11.76  \\
    
    Support Vector Machine & 0.81 & 0.76 & 8.87 & 0.37 & 120.40 & 10.97  \\
    
    Neural Network & \textbf{0.90} & \textbf{0.88} & 5.78 & - & \textbf{61.58} & \textbf{7.85}  \\
  \bottomrule
\end{tabular}
\end{table*}

\FloatBarrier

\section{Conclusion}\label{sec:sec4}
In this study, we explored the prior universities of professors and students majoring in computer science at elite institutions in North America. We examined how an applicant's educational background influences admission decisions. Based on our findings, we observed that the selectivity of a university positively correlates with the number of its students admitted to top universities. Additionally, we discovered that a majority of professors themselves have obtained their doctoral degrees from these top 25 universities. Furthermore, we demonstrated a preference among advisors to admit students who have completed their bachelor's degree at the same university. By analyzing the network of prior universities, we identified distinct communities and observed that Iranian students tend to favor Canadian institutions, while Indian and Chinese students often choose American universities. Moreover, we highlighted a growing trend of students applying directly to doctoral programs without pursuing a master's degree, particularly among women. Finally, we demonstrated that neural networks outperform other regression techniques in predicting the h-index of professors, achieving an RMSE of 7.85.

The results of this paper can provide valuable insights for computer science students planning to pursue higher education at prestigious universities in the USA or Canada. Additionally, admissions committees at top-tier universities can leverage our findings to foster a more equitable admission system and promote diversity and inclusivity among their student 

\bibliographystyle{IEEEtran} 
\bibliography{IEEEabrv, refs} 

\begin{thebibliography}{10}
\providecommand{\url}[1]{#1}
\csname url@samestyle\endcsname
\providecommand{\newblock}{\relax}
\providecommand{\bibinfo}[2]{#2}
\providecommand{\BIBentrySTDinterwordspacing}{\spaceskip=0pt\relax}
\providecommand{\BIBentryALTinterwordstretchfactor}{4}
\providecommand{\BIBentryALTinterwordspacing}{\spaceskip=\fontdimen2\font plus
\BIBentryALTinterwordstretchfactor\fontdimen3\font minus \fontdimen4\font\relax}
\providecommand{\BIBforeignlanguage}[2]{{%
\expandafter\ifx\csname l@#1\endcsname\relax
\typeout{** WARNING: IEEEtran.bst: No hyphenation pattern has been}%
\typeout{** loaded for the language `#1'. Using the pattern for}%
\typeout{** the default language instead.}%
\else
\language=\csname l@#1\endcsname
\fi
#2}}
\providecommand{\BIBdecl}{\relax}
\BIBdecl

\bibitem{she2013international}
Q.~She and T.~Wotherspoon, ``International student mobility and highly skilled migration: A comparative study of canada, the united states, and the united kingdom,'' \emph{SpringerPlus}, vol.~2, no.~1, pp. 1--14, 2013.

\bibitem{ghai2015analysis}
B.~Ghai, ``Analysis \& prediction of american graduate admissions process,'' \emph{Department of Computer Science, Stony Brook University, Stony Brook, New York}, 2015.

\bibitem{slonim2008outlook}
J.~Slonim, S.~Scully, and M.~McAllister, ``Outlook on enrolments in computer science in canadian universities,'' \emph{Information and Communications Technology Council, Government of Canada. Retrieved from http://www. shrc. ca/en/Default. aspx}, 2008.

\bibitem{hendler2008avoiding}
J.~Hendler, ``Avoiding another ai winter,'' \emph{IEEE Intelligent Systems}, vol.~23, no.~02, pp. 2--4, 2008.

\bibitem{stepanova2021hiring}
A.~Stepanova, A.~Weaver, J.~Lahey, G.~Alexander, and T.~Hammond, ``Hiring cs graduates: What we learned from employers,'' \emph{ACM Transactions on Computing Education (TOCE)}, vol.~22, no.~1, pp. 1--20, 2021.

\bibitem{sade2019factors}
M.~Säde, R.~Suviste, P.~Luik, E.~Tõnisson, and M.~Lepp, ``Factors that influence students' motivation and perception of studying computer science,'' in \emph{Proceedings of the 50th ACM Technical Symposium on Computer Science Education}, 2019, pp. 873--878.

\bibitem{wendler2012pathways}
C.~Wendler, B.~Bridgeman, R.~Markle, F.~Cline, N.~Bell, P.~McAllister, and J.~Kent, ``Pathways through graduate school and into careers,'' Educational Testing Service, Tech. Rep., 2012.

\bibitem{kalhor2023diversity}
G.~Kalhor, T.~Zeraati, and B.~Bahrak, ``Diversity dilemmas: Uncovering gender and nationality biases in graduate admissions across top north american computer science programs,'' \emph{EPJ Data Science}, vol.~12, no.~1, pp. 1--23, 2023.

\bibitem{newkirkturner2022do}
B.~L. Newkirk-Turner and T.~K. Hudson, ``Do no harm: Graduate admissions letters of recommendation and unconscious bias,'' \emph{Perspectives of the ASHA Special Interest Groups}, vol.~7, no.~2, pp. 463--475, 2022.

\bibitem{kanojia2017statement}
D.~Kanojia, N.~Wani, and P.~Bhattacharyya, ``Is your statement purposeless? predicting computer science graduation admission acceptance based on statement of purpose,'' in \emph{Proceedings of the 14th International Conference on Natural Language Processing (ICON-2017)}, December 2017, pp. 141--145.

\bibitem{raghunathan2010demystifying}
K.~Raghunathan, ``Demystifying the american graduate admissions process,'' \emph{StudyMode. com}, 2010.

\bibitem{michel2019graduate}
R.~S. Michel, V.~Belur, B.~Naemi, and H.~J. Kell, ``Graduate admissions practices: A targeted review of the literature,'' \emph{ETS research report series}, vol. 2019, no.~1, pp. 1--18, 2019.

\bibitem{posselt2014toward}
J.~R. Posselt, ``Toward inclusive excellence in graduate education: Constructing merit and diversity in phd admissions,'' \emph{American Journal of Education}, vol. 120, no.~4, pp. 481--514, 2014.

\bibitem{lang1987stratification}
D.~Lang, ``Stratification and prestige hierarchies in graduate and professional education,'' \emph{Sociological Inquiry}, vol.~57, no.~1, pp. 12--31, 1987.

\bibitem{zhang2005advance}
L.~Zhang, ``Advance to graduate education: The effect of college quality and undergraduate majors,'' \emph{The review of higher education}, vol.~28, no.~3, pp. 313--338, 2005.

\bibitem{McCarty2013predicting}
C.~McCarty, J.~W. Jawitz, A.~Hopkins, and A.~Goldman, ``Predicting author h-index using characteristics of the co-author network,'' \emph{Scientometrics}, vol.~96, no.~2, pp. 467--483, 2013.

\bibitem{ayaz2018predicting}
S.~Ayaz, N.~Masood, and M.~A. Islam, ``Predicting scientific impact based on h-index,'' \emph{Scientometrics}, vol. 114, pp. 993--1010, 2018.

\bibitem{ibanez2011predicting}
A.~Ib{\'a}{\~n}ez, P.~Larra{\~n}aga, and C.~Bielza, ``Predicting the h-index with cost-sensitive naive bayes,'' in \emph{2011 11th International Conference on Intelligent Systems Design and Applications}.\hskip 1em plus 0.5em minus 0.4em\relax IEEE, 2011, pp. 599--604.

\bibitem{kendall1938new}
M.~G. Kendall, ``A new measure of rank correlation,'' \emph{Biometrika}, vol.~30, no. 1/2, pp. 81--93, 1938.

\bibitem{crane1970academic}
D.~Crane, ``The academic marketplace revisited: A study of faculty mobility using the cartter ratings,'' \emph{American journal of Sociology}, vol.~75, no.~6, pp. 953--964, 1970.

\bibitem{bedeian2010doctoral}
A.~G. Bedeian, D.~E. Cavazos, J.~G. Hunt, and L.~R. Jauch, ``Doctoral degree prestige and the academic marketplace: A study of career mobility within the management discipline,'' \emph{Academy of Management Learning \& Education}, vol.~9, no.~1, pp. 11--25, 2010.

\bibitem{serrano2009extracting}
M.~{\'A}. Serrano, M.~Bogun{\'a}, and A.~Vespignani, ``Extracting the multiscale backbone of complex weighted networks,'' \emph{Proceedings of the national academy of sciences}, vol. 106, no.~16, pp. 6483--6488, 2009.

\bibitem{blondel2008fast}
V.~D. Blondel, J.-L. Guillaume, R.~Lambiotte, and E.~Lefebvre, ``Fast unfolding of communities in large networks,'' \emph{Journal of statistical mechanics: theory and experiment}, vol. 2008, no.~10, p. P10008, 2008.

\bibitem{woroby2015immigration}
T.~Woroby, ``Immigration reform in canada and the united states: How dramatic, how different?'' \emph{American Review of Canadian Studies}, vol.~45, no.~4, pp. 430--450, 2015.

\bibitem{tshibaka2018understanding}
A.~S. Tshibaka, ``Understanding international students from asia in american universities: Learning and living globalization,'' \emph{Journal of International Students}, vol.~8, no.~4, pp. 1941--1943, 2018.

\bibitem{kleinberg1999authoritative}
J.~M. Kleinberg, ``Authoritative sources in a hyperlinked environment,'' \emph{Journal of the ACM (JACM)}, vol.~46, no.~5, pp. 604--632, 1999.

\bibitem{freeman2002centrality}
L.~C. Freeman, ``Centrality in social networks: Conceptual clarification,'' in \emph{Social Network: Critical Concepts in Sociology}.\hskip 1em plus 0.5em minus 0.4em\relax London: Routledge, 2002, vol.~1, pp. 238--263.

\bibitem{pottinger1999choosing}
R.~Pottinger, ``Choosing a ph. d. program in computer science,'' \emph{XRDS: Crossroads, The ACM Magazine for Students}, vol.~6, no.~1, pp. 6--13, 1999.

\bibitem{diciccio1996bootstrap}
T.~J. DiCiccio and B.~Efron, ``Bootstrap confidence intervals,'' \emph{Statistical science}, vol.~11, no.~3, pp. 189--228, 1996.

\bibitem{pedregosa2011scikit}
F.~Pedregosa, G.~Varoquaux, A.~Gramfort, V.~Michel, B.~Thirion, O.~Grisel, M.~Blondel, P.~Prettenhofer, R.~Weiss, V.~Dubourg \emph{et~al.}, ``Scikit-learn: Machine learning in python. the journal of machine learning research 12,'' 2011.

\end{thebibliography}

\end{document}